\documentclass[a4paper,11pt]{article}
\usepackage{scrlfile}
\PreventPackageFromLoading{natbib}
\usepackage{pos}
\usepackage{graphicx}
\usepackage{amsmath}
\usepackage{booktabs}
\usepackage{microtype}
\usepackage{enumitem}
\usepackage{floatrow}
\newfloatcommand{capbtabbox}{table}[][\FBwidth]
\usepackage[compact]{titlesec}

\usepackage{lineno}
\usepackage[backend=biber,style=phys,sorting=none]{biblatex}
\title{A new AI-based Full Event Interpretation model for flavour physics}
\ShortTitle{GNN-based FEI for flavour physics}

\author[a]{Matteo Barbieri}
\author[b]{Marta Calvi}
\author*[a,c]{Elena Graverini}
\author[d]{Rafael Silva Coutinho}
\author[e]{Felipe Luan Souza de Almeida}

\affiliation[a]{Universit\`a di Pisa and INFN Sezione di Pisa, Pisa, Italy}
\affiliation[b]{Università degli studi Milano-Bicocca and INFN Sezione di Milano-Bicocca, Milano, Italy}
\affiliation[c]{\'{E}cole Polytechnique F\'{e}d\'{e}rale, Lausanne, Switzerland}
\affiliation[d]{Centro Brasileiro de Pesquisas Físicas, Rio de Janeiro, Brazil}
\affiliation[e]{Universitat de Barcelona, Barcelona, Spain}
\emailAdd{elena.graverini@cern.ch}

\abstract{The increasing event complexity and data rates foreseen for future LHC coinditions motivate a transition from selecting proton-proton ($pp$) collision events by reconstructing predefined signal candidates to constructing a reduced data stream by interpreting the relevant event features in a signal-agnostic way. We explore the capabilities of deep-learning-based architectures on heavy-flavour physics using LHCb as a case study. Graph-based full-event interpretation techniques address the problem by representing visible products of $pp$ collisions as nodes of a graph, and reconstructing heavy-hadron decay chains hierarchically. A new neutral-particle-aware architecture extends the input data from charged tracks to calorimeter clusters, and uses metric learning to clusterize particles from the same decay chain in a common latent space. Tests on inclusive simulated $pp\to b\bar b$ events show that this approach retains its ability to reconstruct decay chains across a broad range of topologies and multiplicities. We discuss the resulting physics opportunities, with a focus on inclusive $b\to s\ell\ell$ measurements and on searches for exotic hadrons.}

\FullConference{43rd International Conference on High Energy Physics (ICHEP 2026)\\
30 July to 5 August, 2026\\
Natal, Brazil\\}

\begin{document}
\maketitle

\section{Introduction}
\noindent
LHCb is a forward spectrometer at the LHC specialised in the study of beauty- and charm-hadron decays. Its very large heavy-flavour samples provide unique opportunities for precision flavour physics and searches for phenomena beyond the Standard Model. At the same time, the increasing luminosity and event complexity expected for the upgraded experiment place progressively increasing stress on the computing infrastructure devoted to event reconstruction. The integrated luminosity has grown from about $9\,\mathrm{fb}^{-1}$ in Runs~1--2 to about $28\,\mathrm{fb}^{-1}$ in Run~3, and, with the planned detector upgrade, it is expected to reach about $300\,\mathrm{fb}^{-1}$ by the end of Run~5. A tenfold increase in the number of $pp$ collisions per LHC bunch crossing is expected, with $\sim$2000 tracks per event~\cite{UII}.
In this environment, the data reduction strategy necessarily moves from persisting events that contains signal to storing only relevant parts of event data. The LHCb Turbo stream used in Run~3 is a relevant example of the latter approach~\cite{Turbo}.

Conventional real-time event selection relies on dedicated selections for exclusive decay modes, requiring new selections for each additional topology. Events passing at least one selection are retained either in full or in reduced form, potentially including misreconstructed candidates and topologically similar backgrounds.
A Deep-learning-based Full Event Interpretation (DFEI) method was proposed as a complementary, signal-agnostic strategy, both at LHCb~\cite{DFEI} and for FCC-$ee$~\cite{Graverini:2026vta}. Instead of searching for one predefined decay, DFEI attempts to identify and hierarchically reconstruct all heavy-hadron decay chains in an event and to discard particles unrelated to them. Reconstructed objects are represented as nodes of a graph, with edges enconding their pairwise relationships. The graph is then processed by a Graph Neural Network (GNN). The network predicts the order of the lowest common ancestor (LCA) for pairs of final-state particles, thereby encoding the hierarchical structure of their common decay tree. This makes it possible to reconstruct several decay chains simultaneously without specifying the decay modes in advance.

\section{A new approach to full event interpretation}
\noindent
The first DFEI prototype used three sequential GNN stages for node pruning, edge pruning and LCA inference~\cite{DFEI}, but suffered from unfavourable latency scaling. A subsequent heterogeneous GNN (HGNN) incorporated tracks and primary $pp$ collision vertices (PVs) as distinct node types, combining graph pruning, decay reconstruction and PV association in a single multi-task model~\cite{HGNN}. Early edge pruning improved inference speed, while decay-chain reconstruction efficiency increased from about 10\% to 40\% on simulated $pp\to b\bar{b}$ events~\cite{HGNN}.

These prototypes used charged tracks only. This is a severe limitation for a general event-interpretation framework: neutral particles are abundant in heavy-hadron decays, and decays containing photons and neutral pions are central to the LHCb physics programme. However, neutral particles do not leave traces in tracking detectors, thus lacking topological trajectories, and are only detected as electromagnetic calorimeter (ECAL) energy clusters. Adding calorimeter objects roughly doubles the number of final-state objects to fit in the graph. A naive fully-connected graph would therefore lead to a combinatorial growth of the number of edges; the topological depth of the decay trees would also increase to values inaccessible to the previous prototypes.

\begin{figure}
	\centering
	\begin{floatrow}
	\ffigbox{%
 	 \includegraphics[width=\linewidth]{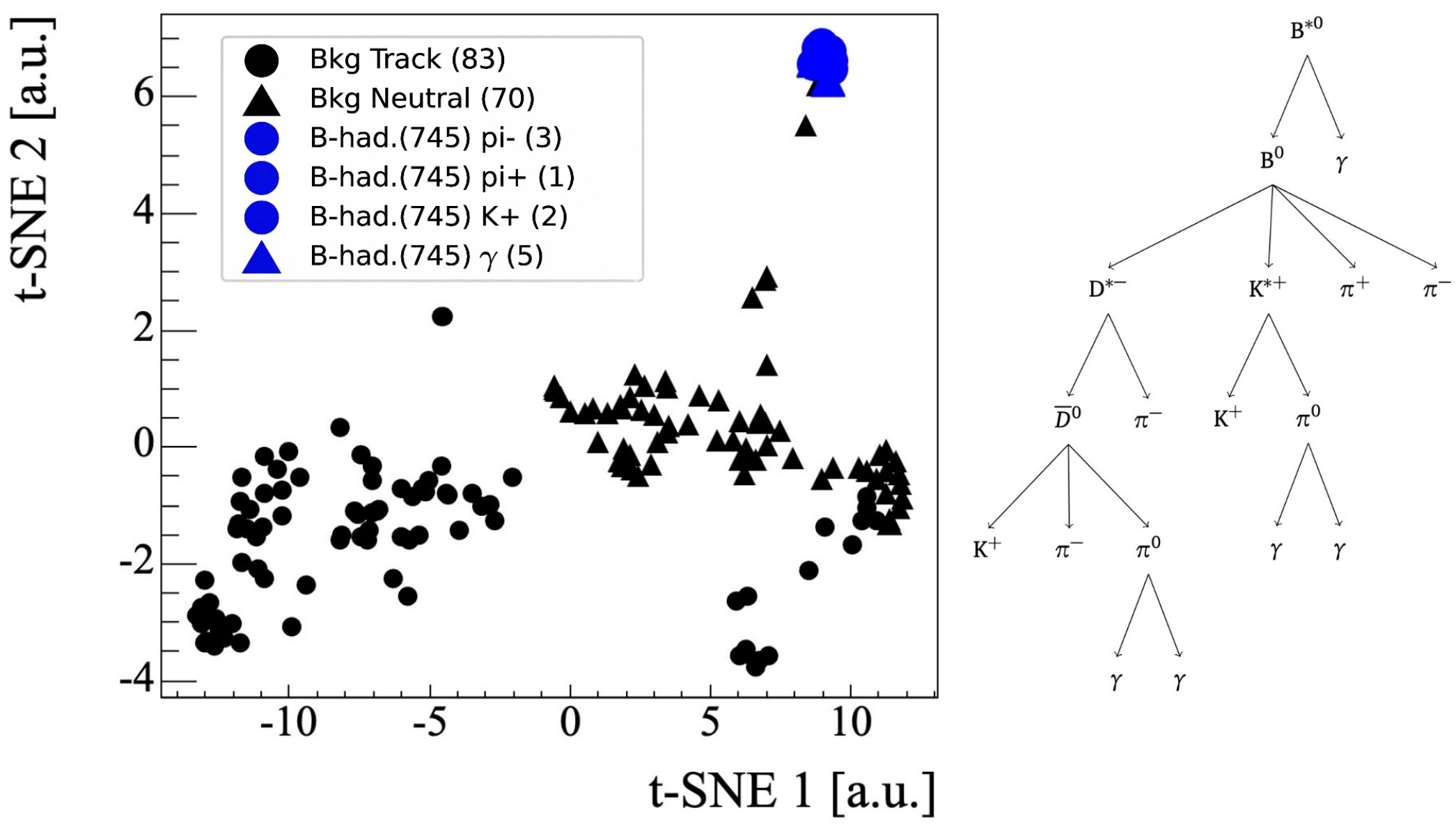}
	}{%
  	\caption{2D-compressed view of the 8D learned representation of an inclusive b-hadron decay event with 11 final state particles and a decay depth of 5.\label{fig:sne}}%
	}
	\capbtabbox{%
\begin{tabular}{lc}
\toprule
Reconstruction category & Fraction [\%] \\
\midrule
Perfect reconstruction & $10.0\pm0.2$ \\
Complete reconstruction & $18.6\pm0.2$ \\
Not isolated (complete) & $11.9\pm0.2$ \\
Not isolated (partial) & $31.8\pm0.3$ \\
Partially reconstructed & $25.6\pm0.3$ \\
Missed & $2.1\pm0.1$ \\
\bottomrule
\end{tabular}
}{%
  \caption{Reconstruction performance of the neutral-aware HGNN on $20\,000$ inclusive $b\bar b$ decays, after training on $200\,000$ inclusive events.}
\label{tab:inclusive}
}
	\end{floatrow}
\end{figure}

A new neutral-aware HGNN architecture addresses this problem. The graph contains three node types: PVs, charged tracks and ECAL clusters. Four classes of edges describe track--track, track--PV, track--cluster and cluster--cluster relations. Track features include kinematic, vertex, impact-parameter and isolation information; ECAL clusters are described by their reconstructed energy, calorimeter position and isolation observables. Pairwise geometric and kinematic quantities provide edge features.
The model is trained with a multi-component loss function,
{
\begin{equation}
 {\cal L}_{\rm total}=\lambda_n {\cal L}_{\rm node}+\lambda_m {\cal L}_{\rm metric}+\lambda_l {\cal L}_{\rm LCA}, \quad\text{with}\quad {\cal L}_{\rm metric}=yd^2 +(1-y)\,\mathrm{max}{(0, m-d)}^2,
\end{equation}}
where the first term handles background rejection, the last term handles hierarchical decay reconstruction, and the novel metric term (with $d$ being an 8-dimensional euclidean distance between two particles, $y$ a flag relating them to the same decay, and $m$ a hyperparameter set to 2.0 via optimisation) allows the model to learn a new representation in which particles originating from the same $b$-hadron decays are mapped closer together, while unrelated particles are separated. An example learned representation is shown in Figure~\ref{fig:sne}, where the particles originating from a particularly complex decay are indicated in blue and all clustered together.

\section{Datasets and tests}\noindent
All datasets used for training and testing the model were obtained using \textsc{Pythia8}~\cite{Pythia} and \textsc{EvtGen}~\cite{EvtGen}. An emulation of the LHCb detector response and reconstruction effects was applied following Appendix~A of Ref.~\cite{DFEI}. We retain only events that fall in the LHCb acceptance, i.e. in the pseudorapidity range $1.9<\eta<4.9$.
We define six classes of inference results:\vspace*{-1mm}
\begin{itemize}[noitemsep]
	\item \textbf{perfect reconstruction:} all decay products are recovered and the decay chain hierarchy is correctly inferred;
	\item \textbf{complete reconstruction:} all decay products are recovered, but the hierarchy is incorrect -- e.g. strong and electromagnetic resonances that decay in place are not recognised as such, due to invariant mass not being an input pairwise feature;
	\item \textbf{not isolated (complete):} all decay products are recovered, alongside some contamination;
	\item \textbf{not isolated (partial):} not all decay products are recovered, and contamination is present;
	\item \textbf{partially reconstructed:} not all decay products are recovered, without contamination;
	\item \textbf{missed:} no decay products are recovered.
\end{itemize}\vspace*{-1mm}
We take as reconstruction efficiency the ratio of events reconstructed either perfectly or completely.

The new model was first tested on simulated exclusive decays containing neutral particles. For $B_s^0\to\phi\gamma$, a reconstruction efficiency of $(89.2\pm0.9)\%$ of the decay chains, with a perfect reconstruction rate of $(86.5\pm1.1)\%$. For $B^\pm\to K^{*\pm}(\to K^\pm\pi^0)\mu^\pm\mu^\mp$, where the intermediate $K^{*\pm}$ meson decays essentially at the production point and is not explicitly resolved by a mass-agnostic algorithm, an efficiency of $(82.3\pm 1.2)\%$, with a perfect reconstruction rate of $(6.5\pm0.9)\%$. 

The model is further tested on a sample of inclusive $pp\to b\bar{b}$ events, containing decay depths up to 7. The neutral-aware HGNN was trained on $2\times10^5$ events and evaluated on $2\times10^4$ independent events. The network manages to group particles from the same decay chain in the learned representation across widely different topologies. Table~\ref{tab:inclusive} gathers the performance obtained. The perfect-reconstruction fraction is approximately twice that of the original GNN prototype, while the complete-reconstruction fraction is approximately three times larger; it is worth noting that the original model only addressed charged particles. These results, while leaving ample space for improvements, establish that inclusive neutral-aware event interpretation is feasible and provide a solid basis for further optimisation and integration in the LHCb processing chain.

\section{Physics opportunities}\noindent
By simultaneously reconstructing multiple heavy-hadron decay chains without assumptions on decay topology or particle masses, DFEI exploits event-wide correlations while suppressing backgrounds. Its inferred decay trees also provide a compact event representation. Potential applications include bremsstrahlung recovery, flavour tagging, CP violation measurements, semileptonic decays, inclusive measurements and searches for exotic hadrons or new physics. Two examples are discussed below.

\paragraph{Inclusive \texorpdfstring{$b\to s\ell\ell$}{b to sll} transitions}\noindent
Rare flavour-changing neutral-current $b\to s\ell\ell$ transitions are particularly sensitive probes of physics beyond the Standard Model. New particles can intervene in the quantum loop or create new Feynman diagrams, with a potential to modify the visible observables associated to such decays. At a hadron collider these decays are normally studied by analysing individual ``exclusive'' decay modes such as $B^+\to K^+\ell^+\ell^-$ or $B^0\to K^{*0}(\to K^+\pi^-)\ell^+\ell^-$. However, the theoretical uncertainties on the hadronic system dynamics grow as decays grow in complexity.
An inclusive reconstruction of the hadronic system would instead provide direct access to the underlying quark-level $b\to s\ell\ell$ transition, while at the same time ensuring the whole experimental data set is effectively used~\cite{Lee:2006gs,Huber:2020vup}.

\begin{figure}
	\centering\includegraphics[width=.45\textwidth]{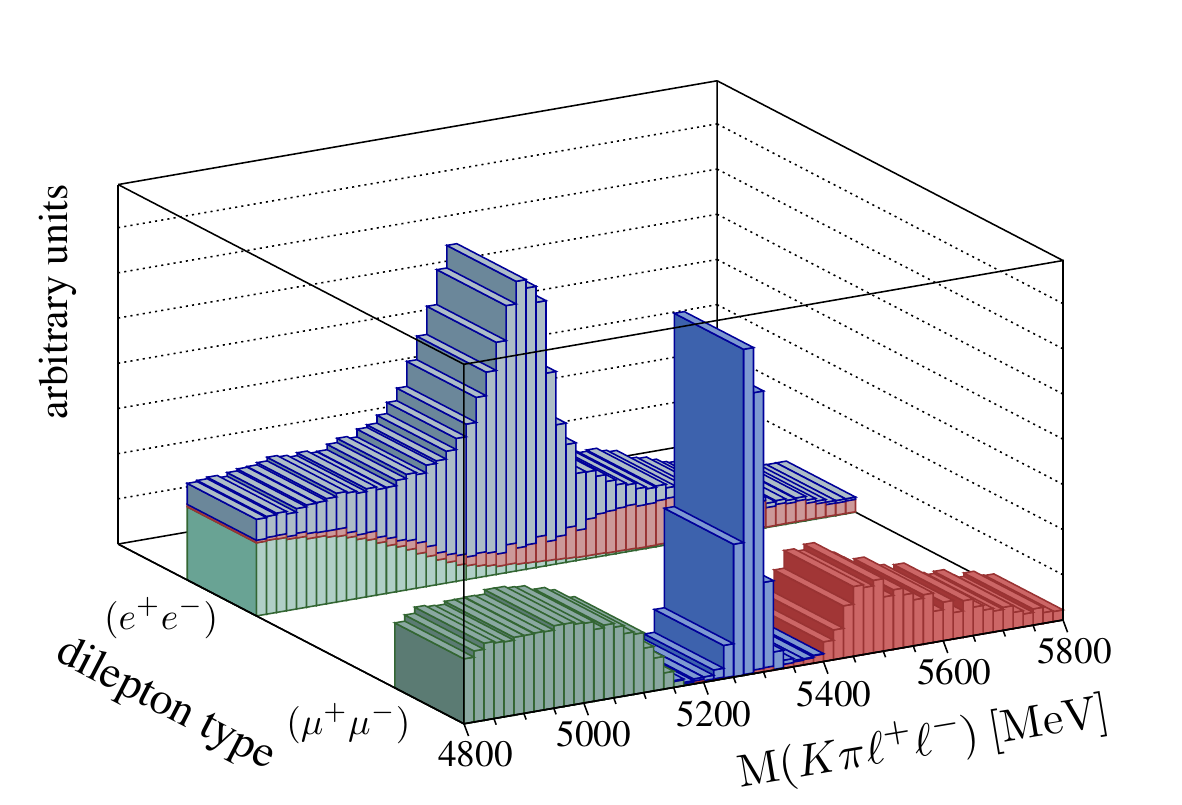}
	\includegraphics[width=.45\textwidth]{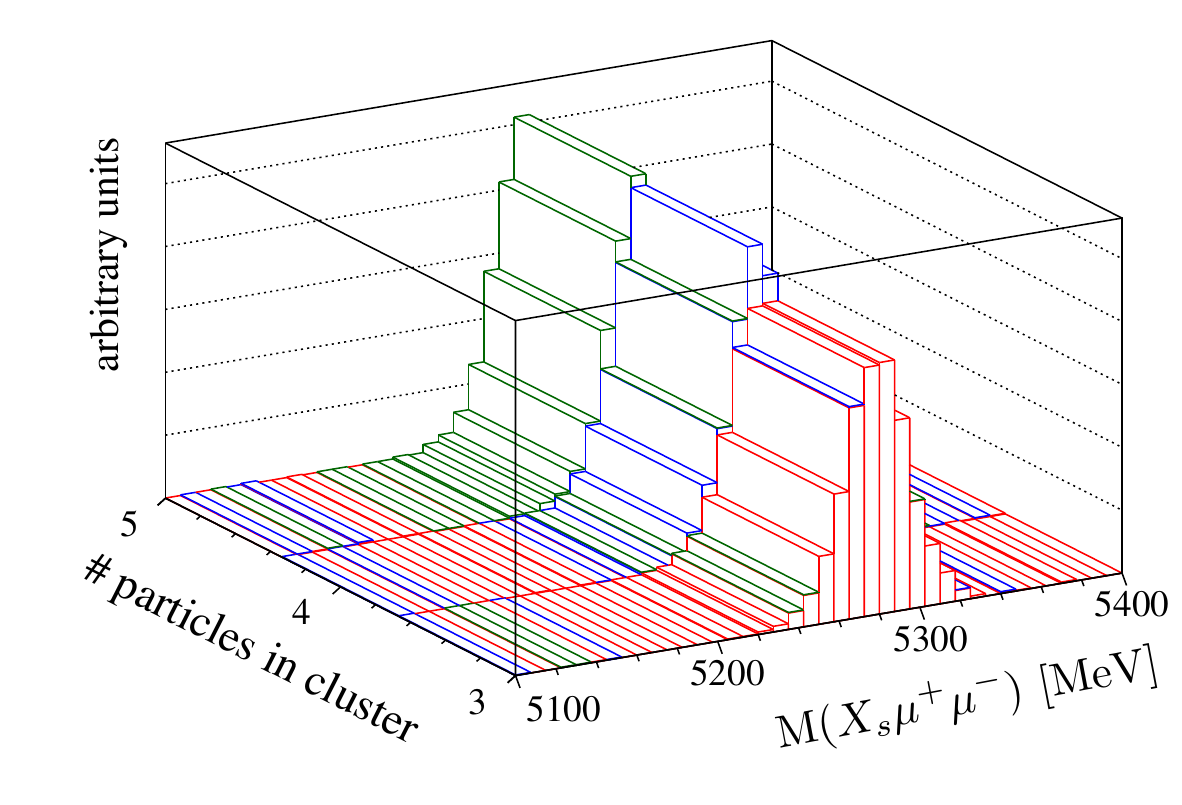}
	\vspace*{-3mm}
	\caption{Invariant-mass spectra for $B\rightarrow K\pi\ell^+\ell^-$ final states, with $\ell=(e,\mu)$, with traditional (left) and graph-based FEI (right) reconstruction. Different colours show different components: $B\rightarrow K\ell^+\ell^-$ (red), $B\rightarrow K\pi\ell^+\ell^-$ (blue), $B\rightarrow K\pi\pi\ell^+\ell^-$ (green).}\label{fig:xsll-trad}
\end{figure}

In traditional signal-based reconstruction algorithms, a decaying hadron is recovered by combining a fixed number of final-state particles. Among the various sources of contamination, similar decays with larger multiplicity, where part of the final state particles are ignored or missed, and decays with lower final-state multiplicity, where an unrelated particle is attached to the candidate, can potentially accumulate close to the signal mass region. These backgrounds are hard to study in traditional analysis and often require to separately measure all of the involved decay channels, combining the results and correlations aftwerwards. An example is shown in Figure~\ref{fig:xsll-trad} (left), where $B^0\to K^+\pi^-\ell^+\ell^-$ signal is polluted by both partially-reconstructed $B^+\to K^+\pi^+\pi^-\ell^+\ell^-$ decays and by combinations of $B^+\to K^+\ell^+\ell^-$ decays with an unrelated pion. In particular, in the $\ell=e$ case, disentangling partially reconstructed $B^+\to K^+\pi^+\pi^- e^+e^-$ decays from radiative tails of the signal has proved to be one of the most significant obstacles at LHCb~\cite{LHCb:2025ilq}.

\begin{figure}[t]
\centering
\hspace*{-6mm}\raisebox{-.5\height}{\includegraphics[width=.4\textwidth]{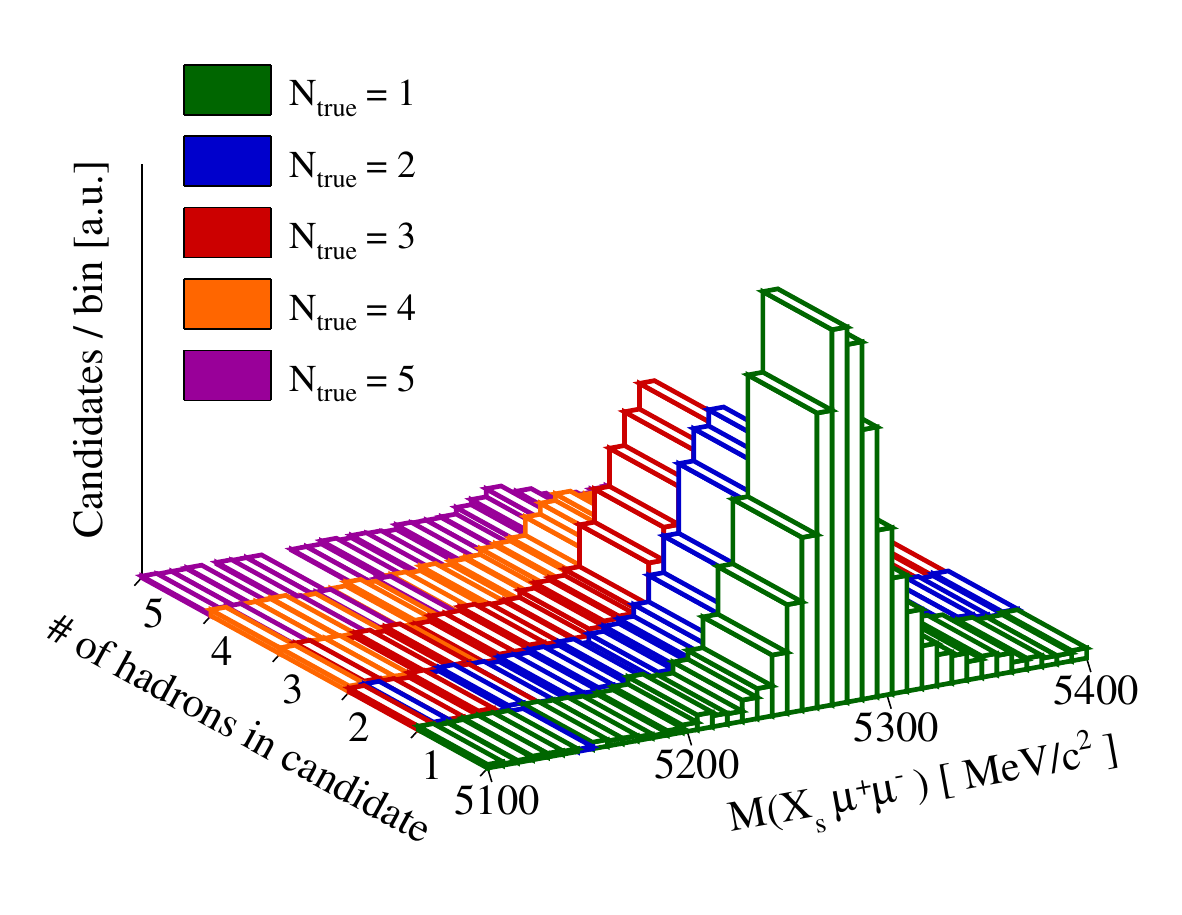}}\hspace*{-2.9mm}
\raisebox{-.5\height}{\includegraphics[width=.4\textwidth]{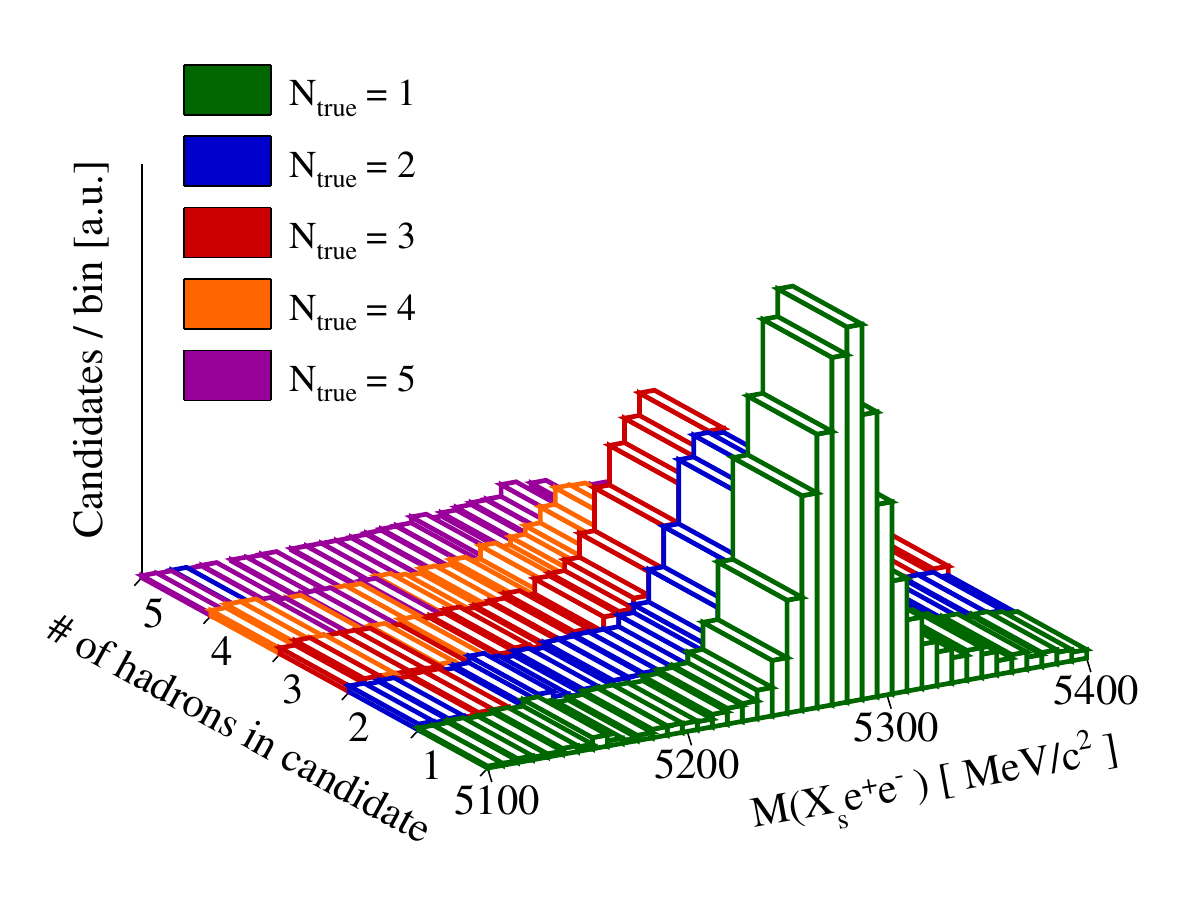}}\hspace*{-2.9mm}
\raisebox{-.5\height}{\includegraphics[width=.3\textwidth]{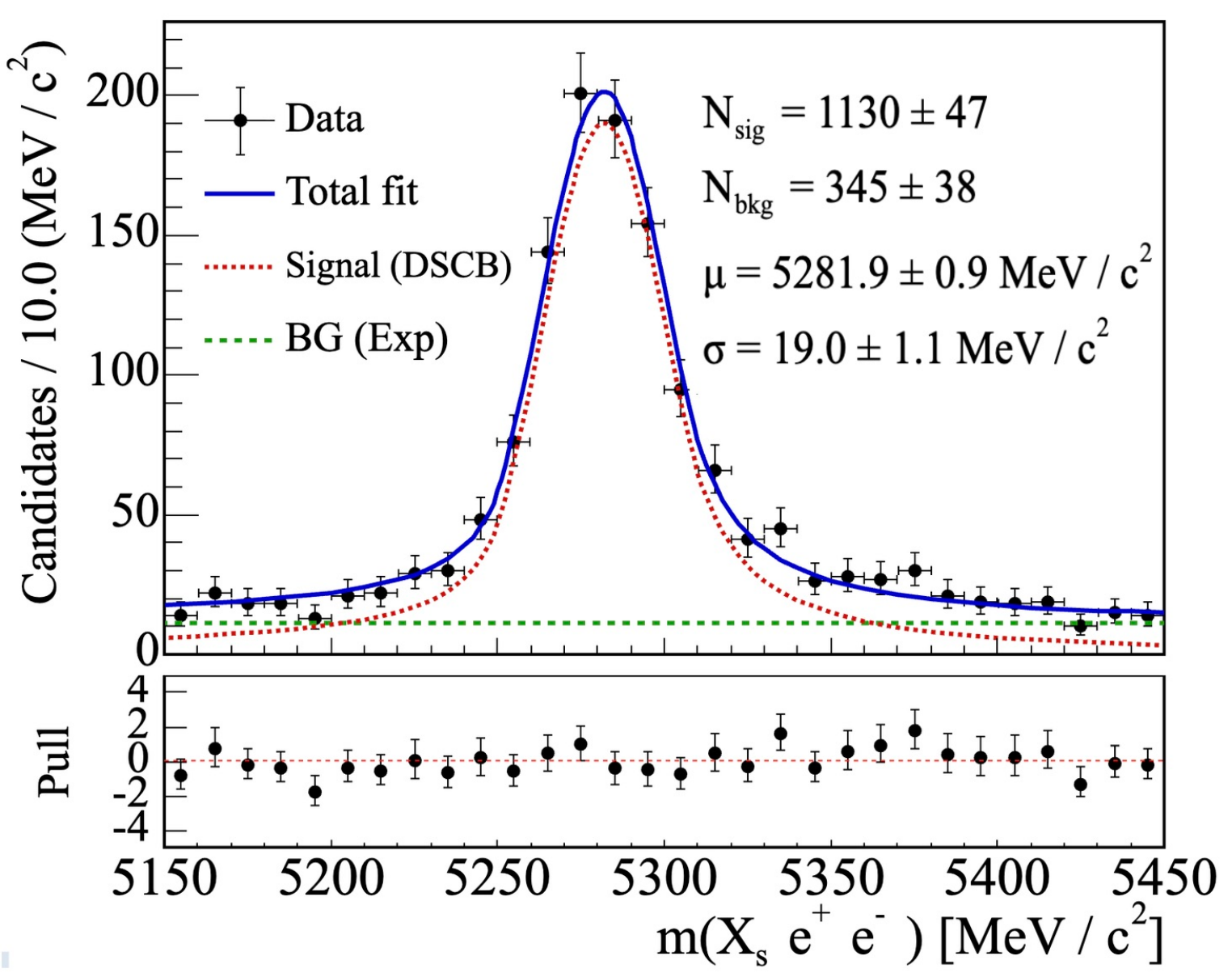}}
\vspace*{-3mm}
\caption{Example application to inclusive $B\to X_s \ell^+\ell^-$ decays. The new HGNN reconstructs candidates across different final-state hadron multiplicities, including neutral hadrons (left, centre); a reconstructed $B\to X_s e^+e^-$ mass peak using all inference categories, including partially/overreconstructed decays (right).}
\label{fig:bsll}
\end{figure}

With the proposed approach, events can be reconstructed all together in a signal-agnostic way and they are natively classified according to their final-state multiplicity, vastly mitigating these kinds of background. This is demonstrated in Figure~\ref{fig:xsll-trad} (right) for the $B^0\to K^+\pi^-\mu^+\mu^-$ case, and in Figure~\ref{fig:bsll} (left, centre) for inclusive $B\to X_s\ell^+\ell^-$ decays, with $\ell=e,\mu$.
The neutral-aware HGNN described above, only trained on inclusive $b\bar{b}$ events, was applied to samples of $25\,000$ $B^{+,0}_{{(s)}}\to X_s e^+e^-$ and $25\,000$ $B^{+,0}_{{(s)}}\to X_s \mu^+\mu^-$ decays, simulated using the EvtGen \texttt{BTOXSLL} and \texttt{BTOSLLBALL} models~\cite{EvtGen}. Despite being trained on a completely different sample, the model achieves efficiencies of $(40.2 \pm 0.3)\%$ and $(37.0 \pm 0.3)\%$ for the muon and electron cases, respectively. The reconstructed $b$-hadron candidates retain information on the number of particles in the decay chain, providing the ingredients for an inclusive analysis in which partially and over-reconstructed contributions can be explicitly used as signal instead of being modelled into complex data-driven background templates.

\paragraph{Mode-agnostic searches for exotic hadrons}
\noindent
The spectroscopy of exotic hadrons provides another natural application. Searches for a new state often require the simultaneous treatment of many possible decay modes, whose branching fractions may be poorly known. A full-event interpretation approach can instead reconstruct candidate decays inclusively to allow a search for a common invariant-mass structure, removing the need to define a list of seeked topologies beforehand.

\begin{figure}\vspace*{-1mm}
    \centering
    \includegraphics[width=0.45\textwidth]{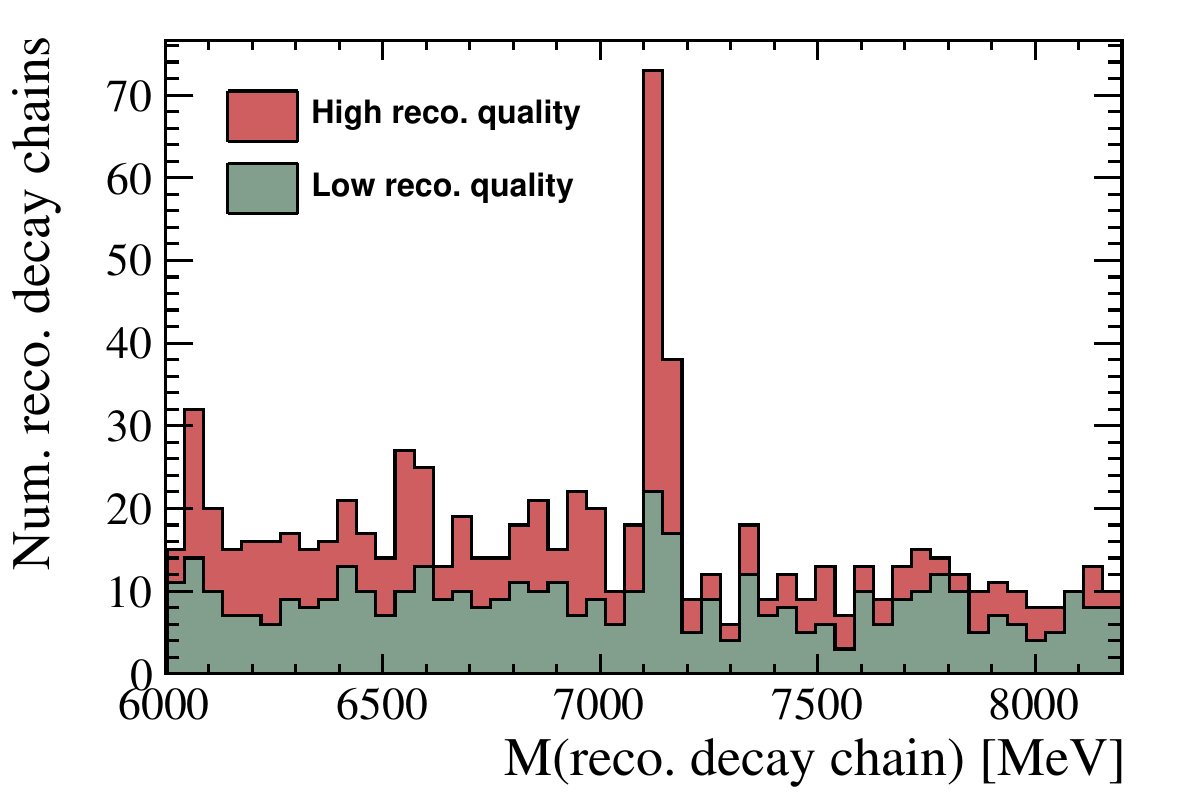}
    \includegraphics[width=0.45\textwidth]{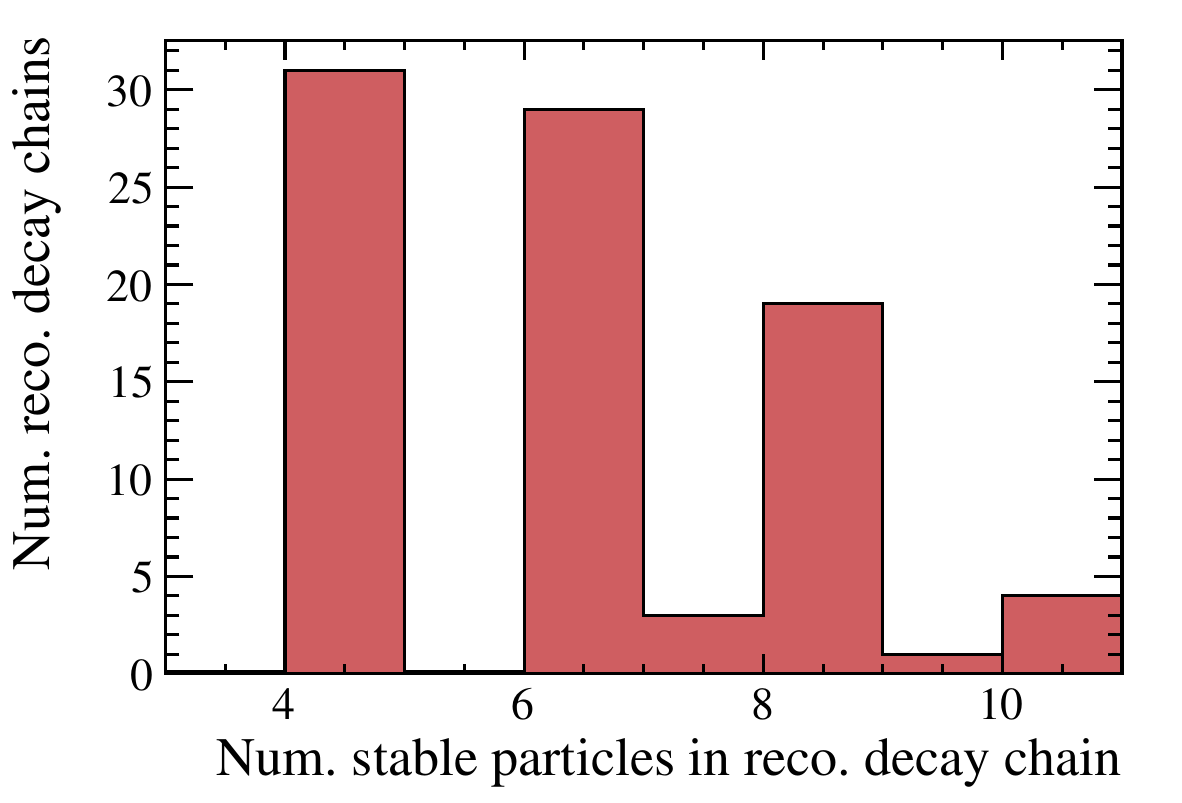}
	\vspace*{-3mm}
    \caption{Combined invariant mass of reconstructed $T_{bc}$ decay chains (left) and final-state multiplicity for events around the $T_{bc}$ mass peak (right).}
    \label{fig:tbc_1}
\end{figure}

This strategy was tested with 41 simulated weak decays of a hypothetical $T_{bc}$ tetraquark, with many different topologies and final-state multiplicities. While the model training sample only contained inclusive $b\bar b$ events with no $T_{bc}$ decays, the inferred decay chains form a clear peak around the generated $T_{bc}$ mass, as shown in Figure~\ref{fig:tbc_1}. This constitutes further proof that a network trained to interpret generic heavy-flavour decays can retain sensitivity to previously unseen topologies.

\section{Conclusions and outlook}
\noindent
We presented a neutral-aware HGNN architecture for signal-agnostic full-event interpretation, extending previous approaches to include calorimeter clusters and enabling the inclusive reconstruction of heavy-hadron decay chains containing neutral particles. A new metric term in the loss function allows to overcome issues related to event complexity. The model performance already supports exploratory studies of inclusive $b\to s\ell\ell$ transitions and mode-agnostic searches for exotic hadrons. 
Future work will address reconstruction quality, inference latency and detector modelling, towards integration into LHCb real-time processing. Ultimately, compact representations of heavy-flavour decays could allow larger physics-relevant samples to be stored, improving sensitivity across the LHCb physics programme.

\printbibliography

\end{document}